# Electron-phonon coupling in crystalline Pentacene films

Richard C. Hatch[1], David L. Huber[2] and Hartmut Höchst[1,*]

[1]Synchrotron Radiation Center, University of Wisconsin-Madison, 3731 Schneider Dr. Stoughton, WI 53589

[2]Department of Physics, University of Wisconsin-Madison, 1150 University Avenue, Madison, WI 53706

The electron-phonon (e-p) interaction in Pentacene (Pn) films grown on Bi(001) was investigated using photoemission spectroscopy. The spectra reveal thermal broadening from which we determine an e-p mass enhancement factor of $\lambda = 0.36 \pm 0.05$ and an effective Einstein energy of $\omega_E = 11 \pm 4$ meV. From $\omega_E$ it is inferred that dominant contributions to the e-p effects observed in ARPES comes from intermolecular vibrations. Based on the experimental data for $\lambda$ we extract an effective Peierls coupling value of $g_{eff} = 0.55$. The e-p coupling narrows the HOMO bandwidth by $15 \pm 8\%$ between 75 K and 300 K.

79.60.Fr, 71.20.Rv, 72.80.Le, 73.50.Gr

For decades theorists have been working to explain the surprisingly high carrier mobilities and temperature dependence of organic semiconductors (OSCs). Pioneering work toward understanding the carrier transport mechanism in OSCs was done by Holstein [1] who studied the influence of electron-phonon (e-p) interaction on mobilities for a model crystal and introduced the concept of a small polaron. More recent work [2] claims that the softness of OSCs allows thermal vibrations to destroy the translational symmetry of the Hamiltonian resulting in charge localization. This theory [2] can be extended to describe charge transport where low energy phonons and librations are able to both localize pure coherent states as well as assist the motion of less coherent ones [3]. Very recently, unified theories were developed that consider both coherent and incoherent contributions to carrier mobility [4, 5]. In these unified theories the e-p coupling plays a major role in both the coherent and incoherent transport. Although theories stress the importance of the e-p coupling in the transport properties of OSCs, surprisingly little work has been performed by experimentalists. We report temperature dependent photoemission data of Pn films and extract the e-p interactions in a fashion similar to studies of various metals and semiconductors [6-8].

The e-p coupling manifests itself in the reduction of the photoemission amplitude as well as in the broadening of photoemission features with increasing temperature [9, 10] from which one can extract the e-p mass enhancement parameter $\lambda$, such that $\mathbf{\textit{m}}^* = \mathbf{\textit{m}}_0(1+\lambda)$ where $\mathbf{\textit{m}}^*$ and $\mathbf{\textit{m}}_0$ are the effective masses with and without e-p coupling respectively.

Crystalline Pn films can be grown on a variety of substrates [11, 12]. We prepared Pn films of at least 100 Å thickness on Bi(001) and studied the temperature dependent photoemission spectra and band dispersions with ARPES. A detailed description of the sample preparation process and of the resulting crystalline film including lattice constants *a* and *b* are reported elsewhere [13]. ARPES experiments were performed on *in situ* grown films at the University of Wisconsin Synchrotron Radiation Center (SRC) where the combined photon and electron energy resolution was $\Delta E \sim 40$ meV. For 15 eV photons (used in all experiments) the electron momentum resolution was $\Delta k_\parallel < 0.03$ Å$^{-1}$.

Since the electronic properties of interest in OSCs are closely related to the upper-most valence bands, we focused our study on the properties of the two highest occupied molecular orbital (HOMO)-derived bands, which will be referred to as the HOMO1 and HOMO2 bands for the lower and higher binding energy bands respectively. Figure 1 shows ARPES spectra of the highest occupied molecular orbital (HOMO) states, HOMO1 and HOMO2, at various locations in *k*-space. Scattering by phonons generally results in a reduction of direct transition probability with increasing temperature [9, 10]. Additionally, most noticeable in Fig. 1(a,d), is a pronounced temperature driven broadening of the photoemission features.

The photoemission features were analyzed following the procedures referred to in Ref. [14]. Our analysis shows thermal broadening of the HOMO1 and HOMO2 bands. The full width at half maximum (FWHM) of these photoemission structures at various locations in *k*-space are shown in Fig. 2. The phonon driven broadening of the photoemission peak can be calculated from the FWHM

$$\Gamma(\omega,T) = \Gamma_0(\omega) + 2\pi\hbar \int_0^{\omega_{max}} d\omega' \alpha^2 F(\omega')[1 - f(\omega - \omega',T) + 2n(\omega',T) + f(\omega + \omega',T)] \quad (1)$$

where $\omega$ is the transition energy, $T$ is the temperature, $\Gamma_0(\omega)$ contains the temperature independent lifetime broadening, $\alpha^2F(\omega)$ is the Eliashberg coupling function, and $f(\omega,T)$ and $n(\omega,T)$ are the Fermi and Bose-Einstein distributions [15]. Two limiting cases for the Eliashberg function come from the Einstein and Debye models for phonon spectra. In the Einstein model, the electrons couple to a single phonon mode of Einstein energy $\omega_E$ and $\alpha^2F(\omega) = \lambda\omega_E\delta(\omega-\omega_E)/2$, while the Debye model considers a distribution of phonon modes with a cutoff at the Debye energy $\omega_D$ and $\alpha^2F(\omega) = \lambda(\omega/\omega_D)^2$ for $\omega<\omega_D$. It should be noted that at high temperatures, independent of the phonon spectrum, the e-p mass enhancement factor $\lambda$ can be extracted from the slope of the linear part of the lifetime broadening using the relation $\lambda = (2\pi k_B)^{-1}d\Gamma/dT$. Fitting Eq. (1) using the Einstein model of $\alpha^2F(\omega)$ to the broadening of the HOMO1 band at $\overline{\Gamma}$ we determined $\lambda = 0.36 \pm 0.05$, and $\omega_E = 11 \pm 4$ meV and $\Gamma_0 = 75 \pm 2$ meV [16]. The temperature induced broadenings of other HOMO features can be described with the same parameters as seen in the inset of Fig. (2) with $\Gamma_0 = 134$, 191 and 223 meV for $\overline{M}_{H1}$, $\overline{M}_{H2}$ and $\overline{\Gamma}_{H2}$ respectively. The $\omega_E$ extracted from our data compares well to the lowest intermolecular modes determined in Refs. [17, 18]. Our value of $\omega_E$ is also in good agreement with an effective average phonon frequency of 13.8 meV extracted [16] from Ref. [19]. A similar fit of Eq. (1) to experimental data using the Debye model of $\alpha^2F(\omega)$ results in $\lambda = 0.37 \pm 0.07$ and $\omega_D = 18 \pm 5$ meV, but may not properly model the excitation of optical phonons in Pn. Both $\omega_E$ and $\omega_D$ lie well below the cut-off for the purely intramolecular vibrations [19], indicating that the e-p interaction observed in ARPES is largely associated with intermolecular modes.

Based on gas phase spectroscopy and quantum mechanical calculations, $\lambda$ can be estimated by considering the density of states at the Fermi level $N(E_F)$ [20, 21]. Assuming $N(E_F)$ is between 2 and 3 eV$^{-1}$ [20] and using the relaxation energies of Ref. [21], one can estimate that $\lambda$ is between 0.2 and 0.3. Earlier, in an effort to explain superconductivity in Pn, a much larger $\lambda = 1.1$ was estimated by normalizing the integral of tunnel microscopy resonances to that of the Pb phonon structure [17].

Using our $\lambda$ it is possible to estimate the effective valence band Peierls coupling value $g_{eff}$. Considering Einstein phonons in the $T=0$ approximation one obtains the relationship $g_{eff} = (\ln(1+\lambda))^{1/2}$. Assuming $\lambda = 0.36$ one obtains a $g_{eff} = 0.55$ which is similar to a theoretically predicted value for Tetracene [22]. This coupling value also seems to support mobility calculations for a model crystal [4], that shows a monotonically decreasing mobility up to room temperature (RT) consistent with experimental data for Pn [23]. Conversely, a stronger coupling value of $g_{eff} \sim 1$ causes the mobility to go through a region of decreasing then increasing mobility up to RT.

The e-p coupling can cause changes in critical point energies and therefore affects the bandwidths. Figure 3(a,b) shows photoemission spectra measured along the electron momentum direction $\overline{\Gamma}$ - $\overline{M}$ (referred to as Pa$_1$) at 75 K and 300 K. The spectra measured at 300 K are broadened but still maintain the characteristic structures as seen in the low temperature spectra. The band dispersions $E(\mathbf{k})$ can be extracted from the data following the procedure outlined in Ref. [13]. Figure 3(c) compares the results of a tight-binding fit considering only the three nearest neighbor interactions. Transfer integrals can be found in the supplementary material [16]. Figure 3(d) shows the 2σ confidence band along with the band dispersions around $\overline{\text{M}}$.

The temperature-induced change in band structure is primarily influenced by the Debye-Waller and Fan contributions to the electron self energy with a typically-neglected, smaller term from the thermal expansion of the lattice [24]. The increase of e-p interaction with temperature causes the bandwidth to narrow. As one can see from Fig. 3(c) the temperature dependent band structure is not homogenous in *k*-space and is most likely due to anisotropic coefficients of thermal expansion [25].

Band narrowing has also been suggested from data of a Pn film grown on graphite [26]. In this work the authors noted a binding energy shift of the center of gravity of their unstructured HOMO emission in the ARPES spectra. It should be noted, however, that while the energy shifts reported in Ref. [26] are probably related to Pn band dispersions, the potentially different Pn polymorph and the lack of spectral features characteristic for emission from the HOMO bands [13, 27] make a direct comparison to the results of this work seem inappropriate [16].

From $E(\boldsymbol{k})$ one can extract the temperature dependence of the effective hole mass tensor $\boldsymbol{m}^*$ which is directly related to carrier mobilities and is given by $\boldsymbol{\mu}_\text{h} = e\tau / \boldsymbol{m}^*$ where $\tau$ is the isotropic scattering time. Figure 4 shows the effective hole mass at the valence band maximum $\overline{\text{M}}$ for 75 K and 300 K. At 75 K the hole mass tensor $\boldsymbol{m}^*$ varies with respect to crystallographic direction from ~14 $m_e$ to ~3 $m_e$ when going from $\boldsymbol{a}$ to $\boldsymbol{b}$.

At 300 K the minimum value is still around *b*, but the maximum increases significantly when approaching the *a* direction. Figure 4 also compares *m\** at 300 K with the effective mass derived from mobility measurements of a Pn single crystal [28]. The e-p interaction not only serves to localize coherent states (such as those around *b*), but also tends to assist the motion of less coherent ones (such as those around *a*). First, the *k*-dependence of the *m\** tensor suggests that the band structure plays a noticeable role in the transport in Pn even at RT. Second, the apparent inconsistency between the band structure-derived effective mass and that from mobility measurements near *a* might be explained by the incoherent contribution to the mobility recently discussed in Ref. [4].

Over the temperature range of 75 K to 300 K the HOMO1 width is reduced from 175 ± 15 meV to 120 ± 25 meV and the HOMO2 band from 250 ± 25 meV to 200 ± 40 meV. In order to compare our data with theoretical predictions of the HOMO bandwidth (HBW), we extract the percentage narrowing of 15 ± 8% between 75 K and 300 K. Our experimentally determined band-narrowing corresponds well with the average value of 12.5% predicted for Pn using extended Hückel theory as well as the INDO/S Hamiltonian [29]. The temperature dependent narrowing in Pn is significantly smaller than those predicted for other oligo-acenes as seen in Fig. 5 where we compare the band narrowing for various molecules. While the smaller molecules lose the majority of their band structure (~ 80% HBW narrowing), the Pn band structure still remains at room temperature and is reduced only by ~15% which is consistent with the expected trend for larger molecules. While the presence of a well defined Pn band structure at RT seems to be in conflict with earlier theoretical models trying to explain the carrier mobilities at elevated temperatures as discussed in Ref. [2], it is incorporated into a more recent theoretical model that unifies the concept of band theory and polaron hopping to predict the mobility of carriers in crystalline OSCs [4]. In this model the band structure contributes to the coherent part of the total carrier mobility at RT.

In conclusion, the e-p coupling in crystalline Pn thin films of high structural quality were analyzed using ARPES. The photoemission spectra reveal thermal broadening from which we determine an e-p mass enhancement factor of $\lambda = 0.36 \pm 0.05$ and an effective Einstein energy of $\omega_E = 11 \pm 4$ meV. From $\omega_E$ it is inferred that the dominant contribution to the e-p effects observed in ARPES comes from intermolecular vibrations. Based on the experimental value for $\lambda$, we can extract an effective Peierls coupling value of $g_{eff} = 0.55$. The e-p coupling leads also to a narrowing of the HBW of $15 \pm 8\%$ between 75 K and 300 K. The effective mass *m*\* shows a strong temperature dependent asymmetry with large changes along the *a* direction of the Pn crystal. Comparison of *m*\* with mobility measurements [28] shows that the band structure plays a noticeable role in the carrier mobility at RT and illustrates the two-fold effect that the e-p interaction has on the transport properties in Pentacene.

We thank Karsten Hannewald for valuable discussions. The SRC, University of Wisconsin-Madison, is supported by the NSF under Award No. DMR-0537588.

[*]Corresponding author: hhochst@wisc.edu

[1] T. Holstein, Annals of Physics **8**, 343 (1959).

[2] A. Troisi, and G. Orlandi, Phys. Rev. Lett. **96**, 086601 (2006).

[3] J. D. Picon, M. N. Bussac, and L. Zuppiroli, Phys. Rev. B **75**, 235106 (2007).

[4] F. Ortmann, F. Bechstedt, and K. Hannewald, Phys. Rev. B **79**, 235206 (2009).

[5] Y. C. Cheng, and R. J. Silbey, J. Chem. Phys. **128**, 114713 (2008).

[6] I. Barke *et al.*, Phys. Rev. Lett. **96**, 216801 (2006).

[7] J. J. Paggel *et al.*, Phys. Rev. Lett. **92**, 186803 (2004).

[8] J. E. Gayone *et al.*, Phys. Rev. Lett. **91**, 127601 (2003).

[9] N. J. Shevchik, Phys. Rev. B **16**, 3428 (1977).

[10] R. Matzdorf, Surf. Sci. Rep. **30**, 153 (1998).

[11] G. E. Thayer *et al.*, Phys. Rev. Lett. **95**, 256106 (2005).

[12] D. Guo *et al.*, Phys. Rev. Lett. **101**, 236103 (2008).

[13] R. C. Hatch, D. L. Huber, and H. Höchst, Phys. Rev. B **80**, 081411 (2009).

[14] B. A. McDougall, T. Balasubramanian, and E. Jensen, Phys. Rev. B **51**, 13891 (1995).

[15] G. Grimvall, in *The Electron-Phonon Interaction in Metals, Selected Topics in Solid State Physics*, edited by E. Wohlfarth (North-Holland, Amsterdam, 1981).

[16] See supplementary material at http://link.aps.org/supplemental/10.1103/PhysRevLett.104.047601 for additional values, discussion, and calculations.

[17] M. Lee *et al.*, Phys. Rev. Lett. **86**, 862 (2001).

[18] K. Fujii *et al.*, Surf. Sci. **601**, 3765 (2007).

[19] A. Girlando *et al.*, Material Science-Poland **22**, 307 (2004).

[20] A. Devos, and M. Lannoo, Phys. Rev. B **58**, 8236 (1998).

[21] V. Coropceanu *et al.*, Phys. Rev. Lett. **89**, 275503 (2002).

[22] K. Hannewald *et al.*, Phys. Rev. B **69**, 075211 (2004).

[23] M. E. Gershenson, V. Podzorov, and A. F. Morpurgo, Rev. Mod. Phys. **78**, 973 (2006).

[24] J. Fraxedas *et al.*, Phys. Rev. B **41**, 10068 (1990).

[25] S. Haas *et al.*, Phys. Rev. B **76**, 205203 (2007).

[26] N. Koch *et al.*, Phys. Rev. Lett. **96**, 156803 (2006).

[27] H. Kakuta *et al.*, Phys. Rev. Lett. **98**, 247601 (2007).

[28] J. Y. Lee, S. Roth, and Y. W. Park, Appl. Phys. Lett. **88**, 252106 (2006).

[29] M. Masino *et al.*, Macromol. Symp. **212**, 375 (2004).

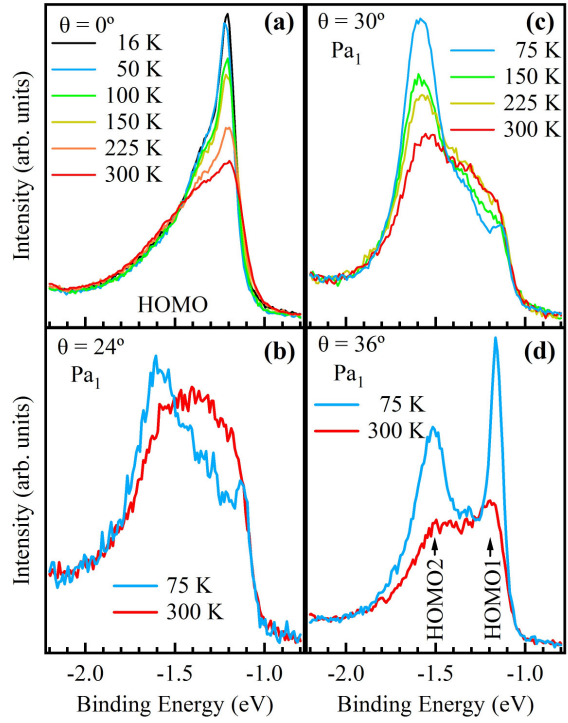

FIG. 1 (Color online). Temperature dependent ARPES spectra of the Pn HOMO region for emission angles θ = 0°, 24°, 30° and 36° shown in (a), (b), (c) and (d) respectively. The emission angle θ is in the $\overline{\Gamma}$ - $\overline{M}$ direction along the reciprocal lattice vector $a^*$–$b^*$ referred to as $Pa_1$.

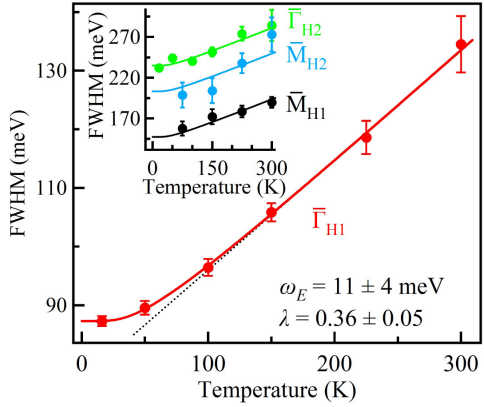

FIG. 2 (Color online). FWHM broadening of HOMO1 photoemission feature H1 (red dots) at $\bar{\Gamma}$. The red line results from a least-square fit of Eq. (1) to the data with $\lambda$ = 0.36±0.05 and an Einstein energy of $\omega_E$ = 11±4 eV. At higher temperatures the broadening is linear with temperature as indicated by the dotted black line. The broadening of the H1 transition at $\bar{M}$ as well as the transitions from the second HOMO band H2 are shown in the inset and can also be fit with the same parameters.

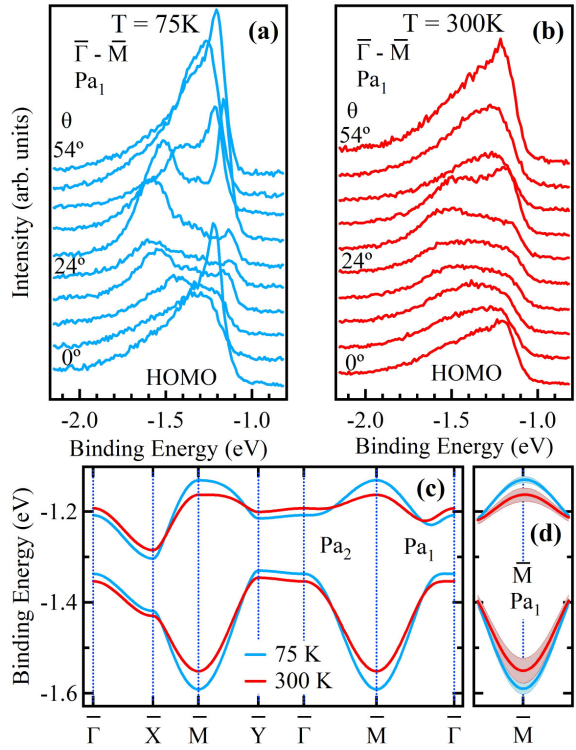

FIG. 3 (Color online). Photoemission spectra of Pn along $Pa_1$ at 75 K and 300 K ((a) and (b) respectively). The band dispersions $E(\mathbf{k})$ resulting from a tight-binding fit to experimental data are shown in (c). $Pa_2$ refers to the direction along the reciprocal lattice vector $\mathbf{a}^*+\mathbf{b}^*$. (d) The $2\sigma$ confidence bands near $\overline{M}$ resulting from the least-squares fit do not overlap.

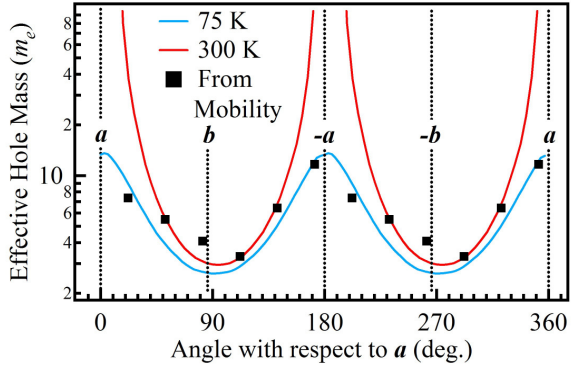

FIG. 4 (Color online). Calculated effective hole mass of Pn at the top of the valence band $\overline{M}$ for 75 K and 300 K plotted in relation to the unit vectors *a* and *b*. Towards the *a* direction, the increase of the hole mass with temperature is significantly more pronounced than in the *b* direction. Black squares are the effective masses calculated from RT mobility measurements from Ref. [28] assuming an isotropic scattering time of $\tau$ = 4.3 fs.

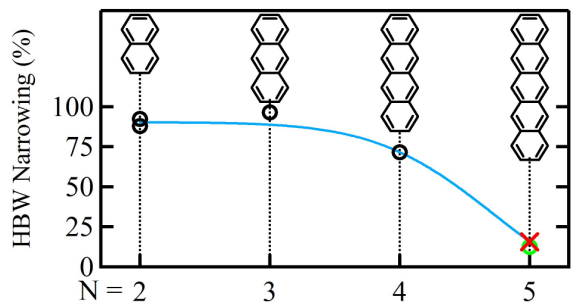

FIG. 5 (Color online). Narrowing of HOMO bandwidth (HBW) between 75 K and 300 K versus the number N of benzene rings for various oligo-acenes. The narrowing is calculated as 100×(HBW(75K) – HBW(300K))/HBW(75K) where HBW is the energy difference between the maximum of HOMO1 to the minimum of HOMO2. Comparison of theory shown with open circles (black Ref. [22] and green (gray) Ref. [29]) with experimental data (red cross). The line through the data serves to guide the eye.

# Electron-phonon coupling in crystalline Pentacene films

Richard C. Hatch[1], David L. Huber[2] and Hartmut Höchst[1,*]

[1]Synchrotron Radiation Center, University of Wisconsin-Madison, 3731 Schneider Dr. Stoughton, WI 53589

[2]Department of Physics, University of Wisconsin-Madison, 1150 University Avenue, Madison, WI 53706

## Supplementary Online Material

**Supplementary TABLE I.** Parameters obtained from fitting a tight binding model to the experimental Pn band structure at 75 K and 300 K. All values are in meV. The subscripts on the transfer integrals define the intermolecular separation. For more information about the tight-binding fit see Ref. [1]

| Fit Parameter | 75 K | 300 K |
|---|---|---|
| $E_0$ | -1317±3 | -1315±8 |
| $\Delta E$ | 114±8 | 145±15 |
| $t_{(a+b)/2}$ | -48±2 | -36±6 |
| $t_{(a-b)/2}$ | 63±2 | 54±9 |
| $t_a$ | 22±3 | 21±5 |

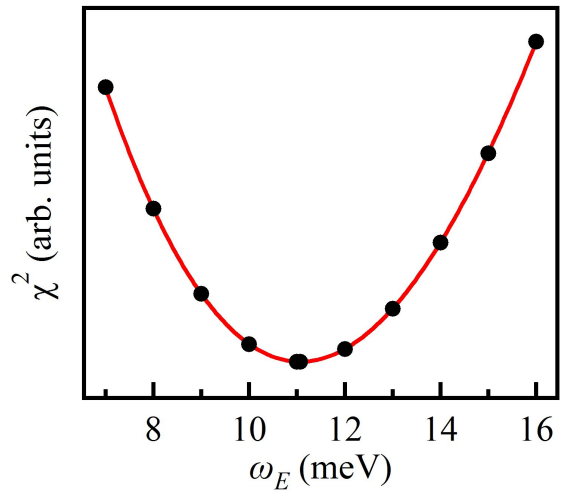

**Supplementary Figure 1.** Resulting $\chi^2$ dependence on Einstein frequency $\omega_E$ obtained by fitting Eq. (1) from text to the thermal broadening of the HOMO1 band at $\bar{\Gamma}$ and by using the Einstein model of $\alpha^2 F(\omega)$. The $\chi^2$ minimum is at 11 meV with a resulting e-p mass enhancement factor of $\lambda = 0.36\pm0.05$.

**Determination of effective average phonon energy**

As explained in Ref. [2] the lattice relaxation energy directly expresses the strength of the Peierls coupling and is defined as $E_j^{LR} = G_j^2 / 2\pi\nu_j$ where $G_j$ is the total Peierls coupling constant of mode *j*, a sum of the modulations of the four main hopping integrals in the **ab** crystal plane. By symmetry, only totally symmetric ($A_g$) phonons are appreciably coupled to the charge carriers. Table 2 in Ref. [2] shows the phonons that are coupled to the charge carriers having an energy less than 203 cm$^{-1}$. Phonons of higher energy are significantly less coupled [2]. From the values listed in Table 2 and from our definition of the effective Einstein energy,

$$\omega_{eff} = \sum_j \frac{G_j}{G_{tot}} \cdot \nu_j = \frac{1}{G_{tot}} \cdot \sum_j G_j \nu_j$$

where $G_{tot}$ is defined as $G_{tot} = \sum_j G_j$ we obtain a value of $\omega_{eff}$ = 13.8 meV. This calculated value for the effective Einstein energy is in excellent agreement with our experimentally determined value of $\omega_E$ = 11 ± 4 meV.

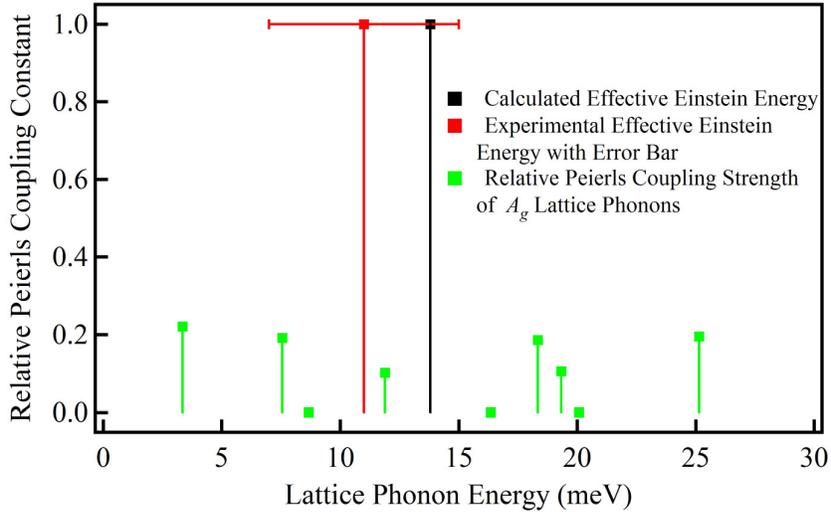

**Supplementary Figure 2. Relative Peierls coupling constants for a given phonon mode energy determined from Ref. [2] (green). Also shown is the calculated and experimentally determined effective Einstein energy (black and red respectively). The calculated and experimentally determined values are, within error bars, in excellent agreement.**

**Determination and discussion of the HOMO bandwidth in Pentacene**

Unlike in PTCDA (Ref. [3]), a solid with one molecule per unit cell, the two inequivalent molecules per Pn unit cell give rise to two HOMO-derived Pn bands which we call HOMO1 and HOMO2. To see how these two bands manifest themselves in photoemission spectra see Fig. 1(d) of the current work and to see how the band dispersions are revealed using angle-resolved photoemission spectroscopy see Ref. [1]. These two bands are predicted by theory [4-8] and have been measured experimentally [1, 9], are separated by ~500 meV and are periodic in symmetry directions. Any determination of the Pn band dispersions and subsequently, the HOMO bandwidth, must consider and distinguish the two bands unlike the work of N. Koch *et al.* [10] where the Pn HOMO feature is treated with a single band.

The definition of the HOMO bandwidth (HBW) is the energy difference between the maximum of HOMO1 and the minimum of HOMO2, and as shown by theory and experiment is ~500 meV [1, 4-9]. In a previous work [1] we determined that the HOMO1 maximum and the HOMO2 minimum coincided at $\overline{M}$. In the current work we report (Fig. 3(d)) our determination of the Pn band structure around $\overline{M}$ at 75 K and 300 K. The 2σ confidence bands (also shown in Fig. 3(d)) are either distinctly separated (for the HOMO1 band) or barely touch (for the HOMO2 band). This error analysis shows that the reported temperature dependent HBW is of statistical significance and we report a HBW narrowing of 15 ± 8%.

*Corresponding Author: hhochst@wisc.edu

[1]   R. C. Hatch, D. L. Huber, and H. Höchst, Phys. Rev. B **80**, 081411 (2009).
[2]   A. Girlando, M. Masino, A. Brillante *et al.*, Materials Science-Poland **22**, 307 (2004).
[3]   G. N. Gavrila, H. Mendez, T. U. Kampen *et al.*, Appl. Phys. Lett. **85**, 4657 (2004).
[4]   G. A. de Wijs, C. C. Mattheus, R. A. de Groot *et al.*, Synth. Met. **139**, 109 (2003).
[5]   H. Yoshida, and N. Sato, Phys. Rev. B **77**, 235205 (2008).
[6]   A. Troisi, and G. Orlandi, J. Phys. Chem. B **109**, 1849 (2005).
[7]   K. Hummer, and C. Ambrosch-Draxl, Phys. Rev. B **72**, 205205 (2005).
[8]   P. Parisse, L. Ottaviano, B. Delley *et al.*, J. Phys.: Condens. Matter **19**, 106209 (10pp) (2007).
[9]   O. Manabu, S. Tadamasa, S. Toshihiro *et al.*, Appl. Phys. Lett. **95**, 123308 (2009).
[10]  N. Koch, A. Vollmer, I. Salzmann *et al.*, Phys. Rev. Lett. **96**, 156803 (2006).